\documentclass{ws-ijmpe}

\begin{document}

\markboth{Ferenc Sikl\'er}{Low $p_T$ Hadronic Physics with CMS}

\catchline{}{}{}{}{}

\title{LOW $\mathrm{P_T}$ HADRONIC PHYSICS WITH CMS}

\author{\footnotesize FERENC SIKL\'ER}

\address{KFKI Research Insitute for Particle and Nuclear Physics \\
Konkoly-Thege Mikl\'os \'ut 29-33, Budapest, H-1121, Hungary \\
sikler@rmki.kfki.hu}

\author{for the CMS Collaboration}

\address{}

\maketitle

\begin{history}
\received{(22 February 2007)}
\revised{(revised date)}
\end{history}

\begin{abstract}

The pixel detector of CMS can be used to reconstruct very low $p_T$ charged
particles down to about 0.1~GeV/$c$. This can be achieved with
high efficiency, good resolution and a negligible fake rate for elementary
collisions. In the case of central PbPb collisions
the fake rate can be kept low for $p_T>0.4$~GeV/$c$. In
addition, the detector can be employed for identification of neutral hadrons
(V0s) and converted photons.

\end{abstract}

\newcommand{\pt}{p_T}
\newcommand{\dedx}{\mathrm{d}E/\mathrm{d}x}

\newcommand{\PKzS}{\ensuremath{\mathrm{K^0_S}}}
\newcommand{\PgL}{\ensuremath{\mathrm{\Lambda}}}
\newcommand{\PagL}{\ensuremath{\mathrm{\overline{\Lambda}}}}

\newcommand{\PKm}{\ensuremath{\mathrm{K^-}}}
\newcommand{\PgXm}{\ensuremath{\mathrm{\Xi^-}}}
\newcommand{\PgOm}{\ensuremath{\mathrm{\Omega^-}}}

\section{Introduction}

The reconstruction of low $\pt$ charged and neutral hadrons (yields, spectra
and correlations) is crucial to characterize the collective properties of the
system produced in nucleus-nucleus collisions at the LHC. In pp collisions,
the measurement of high $\pt$ observables also requires good understanding of
the characteristics of the underlying event and backgrounds which are
dominated by soft $p_T$ spectra~\cite{Revol:2002gn}. 

In CMS, the measurement of charged particle trajectories is achieved
primarily using the silicon tracker with both pixels and strips, embedded in
a 4 T magnetic field, and with geometric coverage over $|\eta|<$ 2.5. The
high granularity silicon pixel tracker consists of three barrel layers (at
about 4, 7 and 11 cm radius) and two endcap disks.  There are about 66
million pixels with an area of 100 $\times$ 150 $\mu\mathrm{m}^2$ and a sensor
thickness of 300 $\mu$m.  The strip part is a combination of single- and
double-sided layers with ten barrel and nine forward layers on each side (9.3
million channels).  The silicon tracker has excellent reconstruction
performance for $\pt > 1$ GeV/$c$: 95\% efficiency for charged hadrons with
high $\pt$, better than 98\% for muons in pp and pA collisions and around
75\% for central PbPb~\cite{Roland:2006kz}.

The reconstruction capabilities at lower $\pt$ are limited by the high
magnetic field and effects of the detector material. In addition,  in central
AA collisions the high occupancy of the silicon strips makes the inclusion of
these strips in charged particle tracking difficult~\cite{Roland:2006kz}.
Using only silicon pixels allows the same analysis to be used for low
multiplicity pp, pA and high multiplicity AA events.

\section{Track reconstruction}

We have developed an improved tracking algorithm which
reconstructs tracks down to 0.1 GeV/$c$, using just the three pixel
layers, with the modified hit triplet finding and cleaning procedures
described here.

\subsection{Modified hit triplet finding}

\begin{figure}
 \centering
  \includegraphics[width=0.43\textwidth]{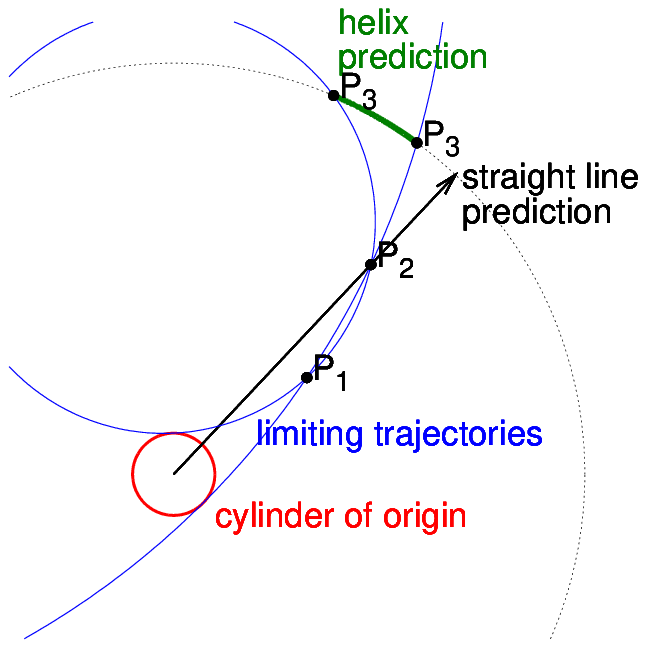}
  \includegraphics[width=0.49\textwidth]{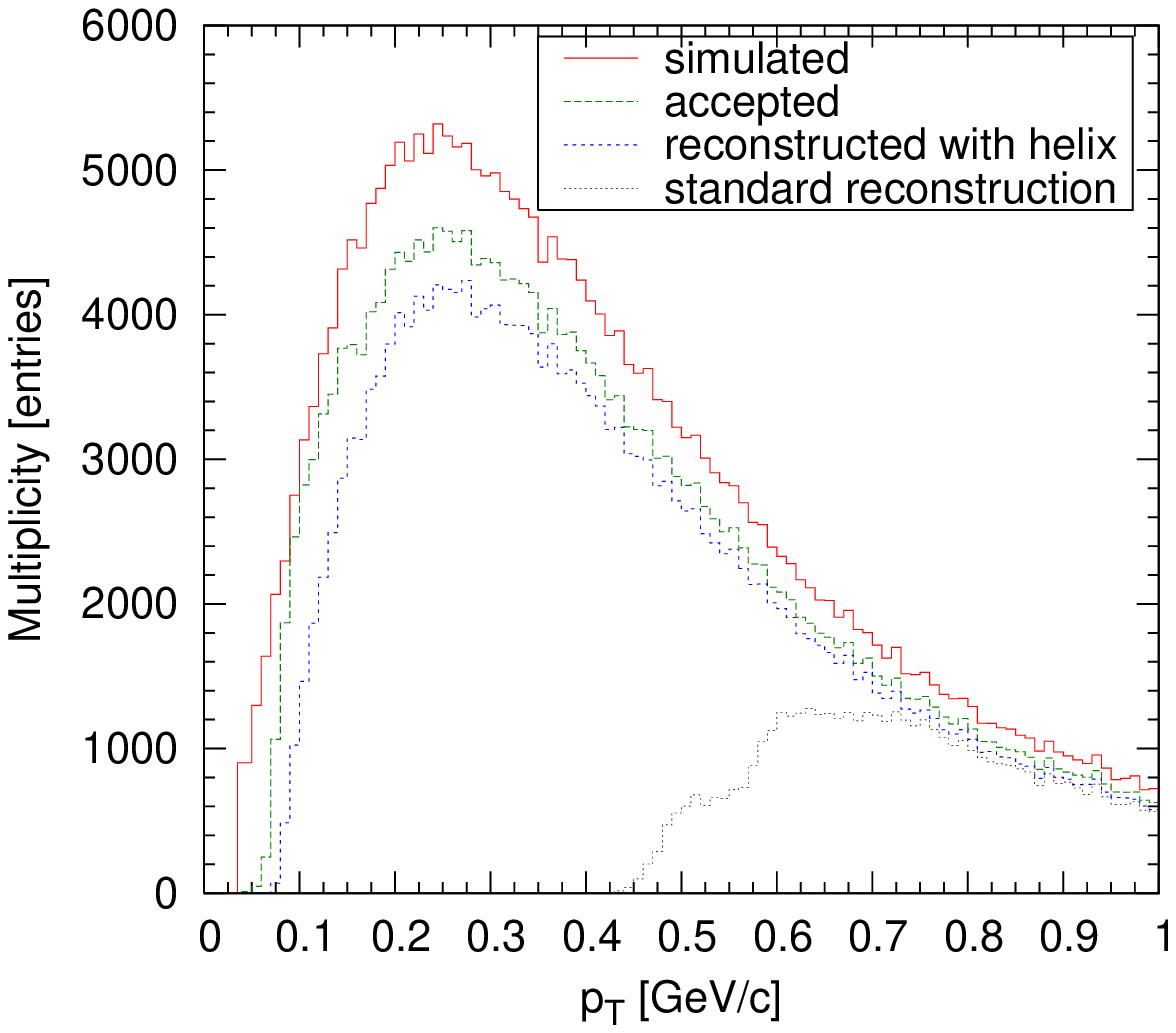}

 \caption{Left: Schematic comparison of the standard straight line prediction
and the new helix prediction for finding the third hit.  Right: Transverse
momentum distributions of the charged particles: simulated (solid red),
accepted (green dashed) and reconstructed, with the standard method (dotted
black) or with the new helix method (dotted blue).} 

 \label{fig:comp_multi}
\end{figure}

The track finding procedure starts by pairing two hits from different layers
(see Fig.~\ref{fig:comp_multi}). During the search for the third hit, the
following requirements must be fulfilled: the track must come from the
cylinder of origin (given by its radius, half-length and position along the
beam-line); the $\pt$ of the track must be above the minimal value $p_{\rm
T,min}$; and the track must be able to reach the layer where the third hit
may be located. In the small volume of the pixel detector the magnetic field
is practically constant and the charged particles propagate on helices. The
projection of a helix or a cylinder onto the transverse plane is a circle.
Each requirement defines a region of allowed track trajectories. They are
enclosed by a pair of limiting circles which can be constructed using simple
geometrical transformations. A third hit candidate is accepted if its
position is within a region which takes into account the expected 
multiple scattering. More details are given in Ref.~\cite{Sikler:lowpt}.

\subsection{Triplet cleaning}

While high $\pt$ tracks are relatively clean, uncorrelated hit clusters
can often be combined to
form fake low $\pt$ tracks.  However, a cluster contains more information than
its position. The geometrical shape of the hit cluster depends on the angle
of incidence of the particle: bigger angles will result in longer clusters.
We can, thus, check whether the measured shape of the
cluster is compatible with the predicted angle of incidence of the track; if
any of the hits in the triplet is not compatible, the triplet is removed
from the list of track candidates.

\subsection{Low $\pt$ tracking results \label{sec:lowpttracking_results}}

The low $p_T$ reconstruction studies are based on 25~000 minimum bias pp
events (generated with Pythia, with the default {\it minimum bias} settings),
reconstructed with the modified hit triplet finding. The algorithm uses the standard CMS
settings, except for a much lower minimum $\pt$ (0.075~GeV/$c$). 

\begin{figure}
 \centering
  \includegraphics[width=0.49\textwidth]{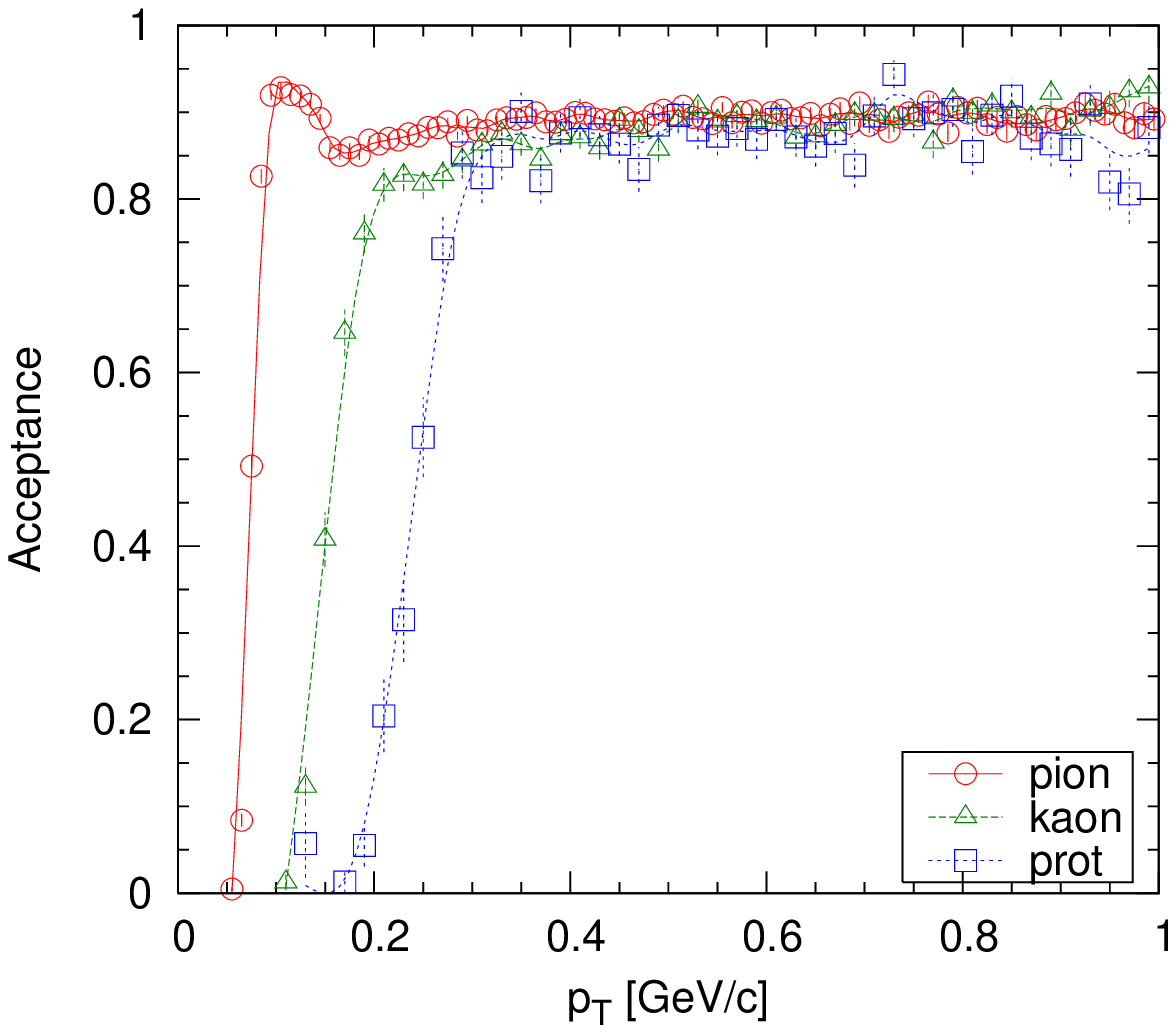}
  \includegraphics[width=0.49\textwidth]{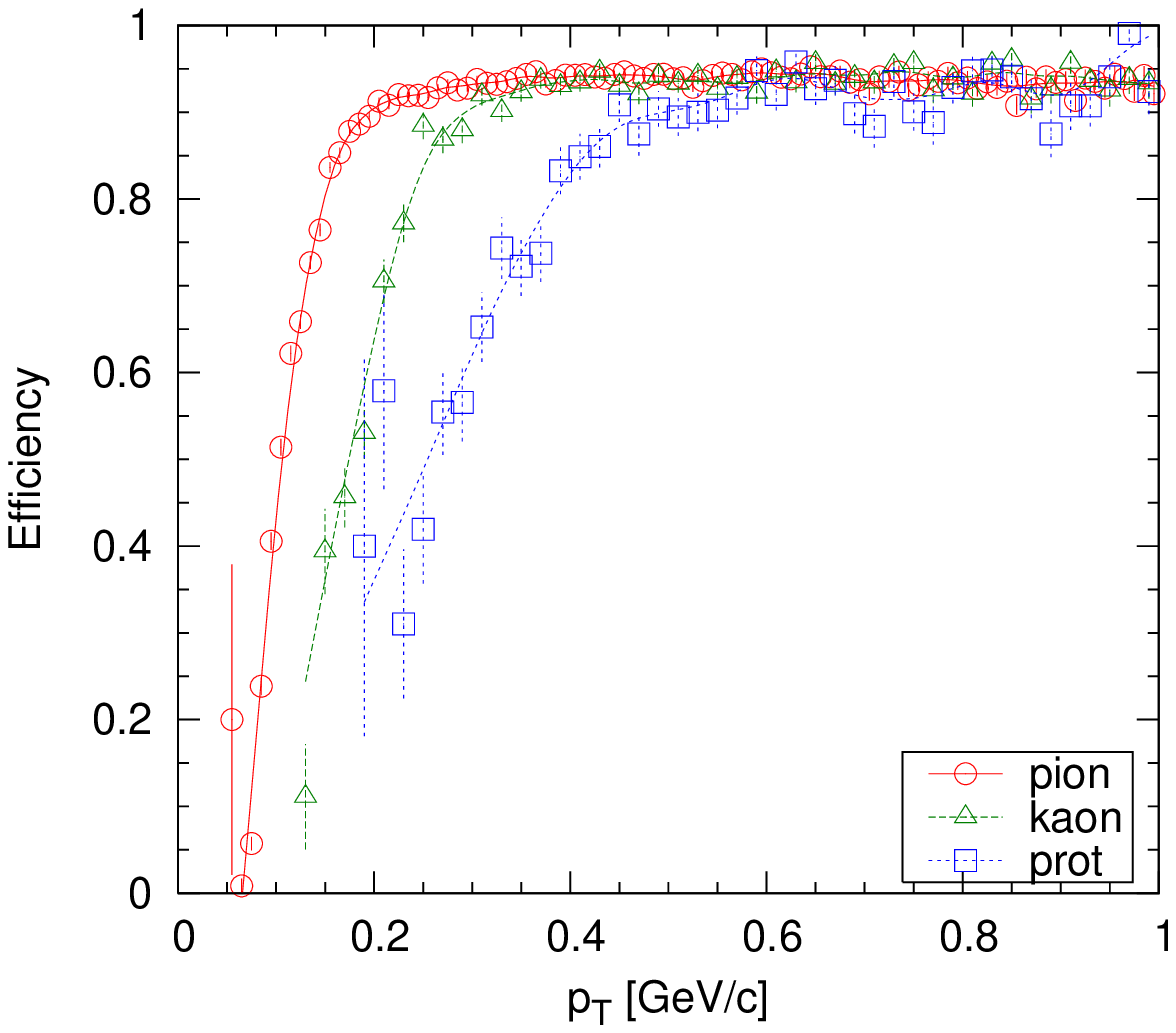}

  \caption{Acceptance (left) and reconstruction efficiency (right) as a function of $\pt$,
for tracks in the range $|\eta|<1$, for pions
(circles), kaons (triangles) and (anti)protons (squares).}

  \label{fig:acceff}
\end{figure}

The acceptances rise sharply with $\pt$ (see Fig.~\ref{fig:acceff}-left), and
become approximately flat above $\pt$ values around 0.1, 0.2 and 0.3~GeV/$c$,
respectively for pions, kaons and protons.  In the range $|\eta|<2$, their
averages are 0.88 (pions), 0.85 (kaons) and 0.84 (protons).
Also the reconstruction efficiencies rise sharply with $\pt$ (see
Fig.~\ref{fig:acceff}-right), and become nearly flat above $\pt$ values
around 0.2, 0.3 and 0.4~GeV/$c$, respectively for pions, kaons and protons.
In the range $|\eta|<1.5$, the corresponding average reconstruction efficiencies are 0.90,
0.90 and 0.86.

Without triplet cleaning, the fake rate is $\sim$~4\% at $\eta \sim 0$ and
reaches 20\% at $|\eta| \sim 2$. With cleaning, the fake rate decreases very
significantly (by a factor of 10), to around 0.5\% and 2\% at $\eta \sim 0$
and $\sim 2$, respectively.  In the range $|\eta|<1$, the fake rate decreases
steeply with $\pt$, being about 4\% at 0.1~GeV/$c$, $\sim$ 1\% at
0.16~GeV/$c$ and at the per mil level for higher $\pt$ values.

\begin{figure}
 \centering 
  \includegraphics[width=0.49\linewidth]{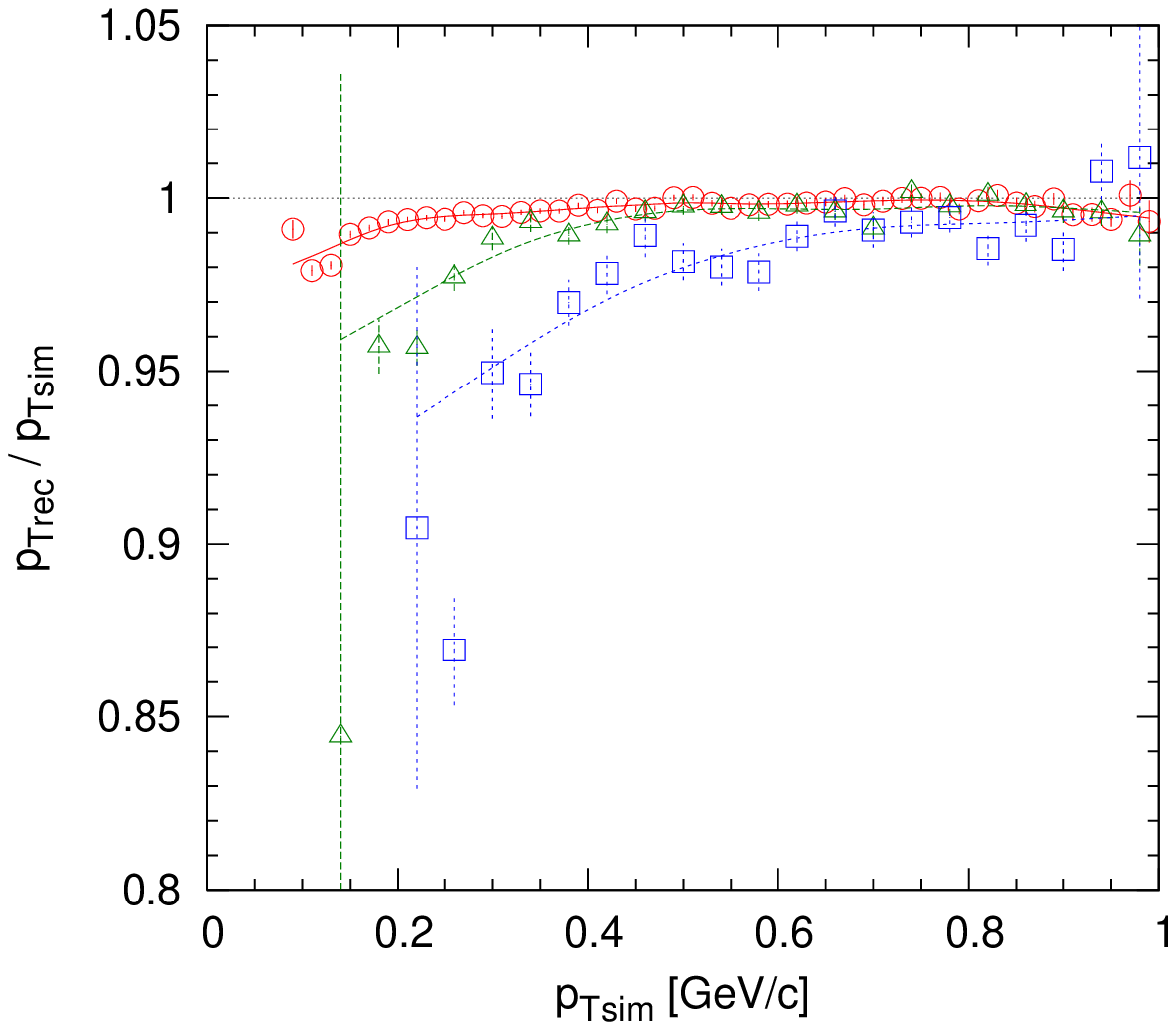}
  \includegraphics[width=0.49\linewidth]{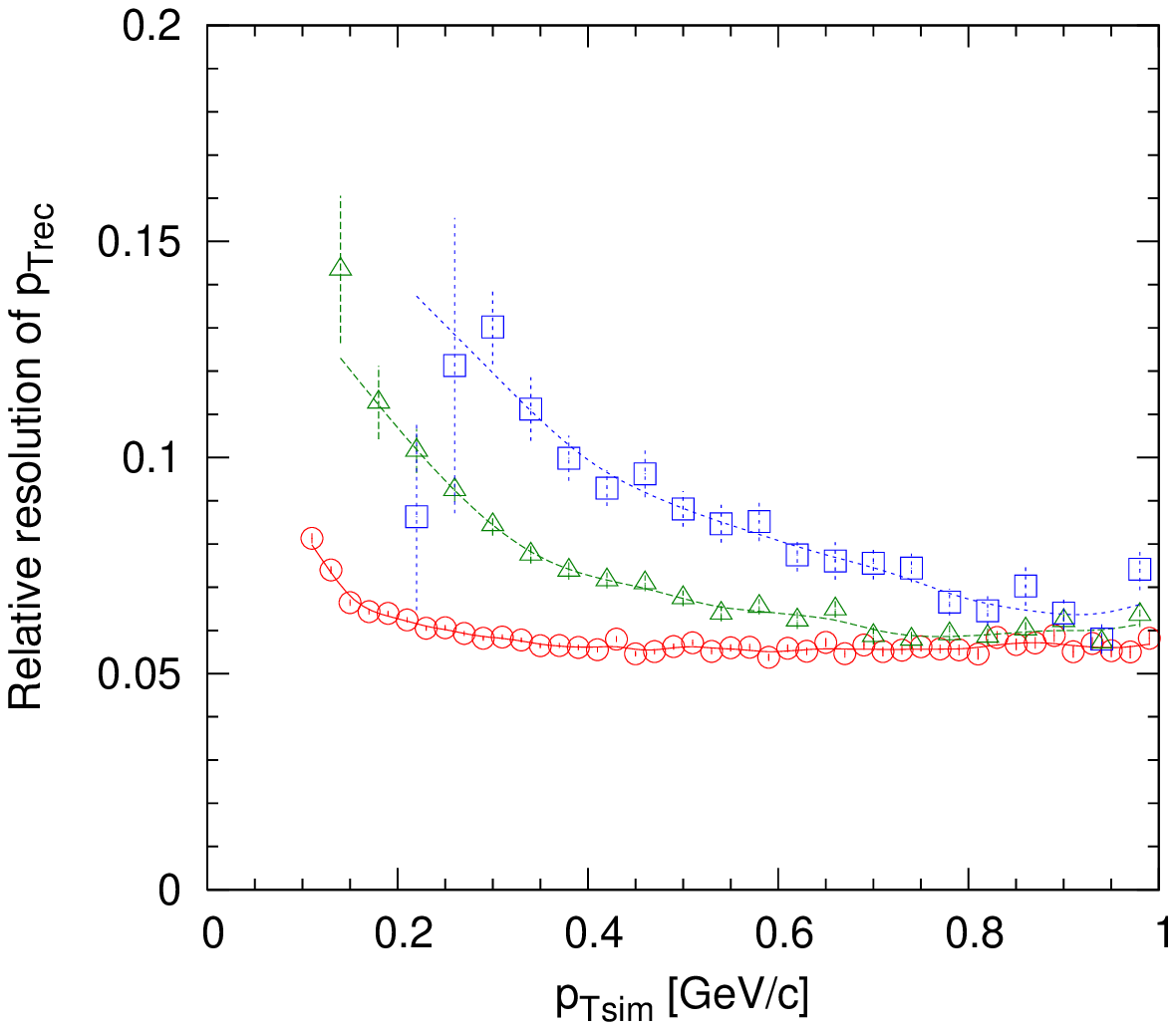}
 \caption{Degradation of the reconstructed $\pt$ as a function of the
 simulated $\pt$, in terms of ``bias'' (left) and resolution (right), 
 for pions (circles), kaons (triangles) and (anti)protons (squares),
 in the range $|\eta|<1$.}
 \label{fig:ptresol}
\end{figure}

Figure~\ref{fig:ptresol}-left shows, as a function of the generated $\pt$ and
separately for pions, kaons and protons, the ratio between the reconstructed
and the simulated $\pt$ (``bias'').  It is seen that the particles generated
at low $\pt$ tend to be reconstructed with a slightly lower $\pt$ value,
because of energy loss effects. This bias is negligible for high $p/m$ values
but is quite significant for low momentum protons (or antiprotons): a
correction of almost 10\% is needed for protons of $\pt \sim 0.2$~GeV/$c$. 
Figure~\ref{fig:ptresol}-right shows how the resolution of the reconstructed
$\pt$ depends on the generated $\pt$.  While at high $\pt$ values the
resolution is $\sim$~6\% for all particles, at low $\pt$ the multiple
scattering and energy straggling effects are more important and lead to
significantly degraded resolutions, in particular for protons.

The performance of the low $\pt$ reconstruction was studied under several running
conditions. These studies are based on 25~000 minimum bias pp events (Pythia
generator) and on 25 central PbPb events (Hydjet generator) with two
multiplicity settings: total particle multiplicities 30\,000 (``central'')
and 15~000 (``mid-central''). In the PbPb case, the primary vertex of the
event was determined first, with good precision, using high $\pt$ tracks.  In
a second step, the cylinder of origin was centered on this vertex, with a
small half-length of 0.1~cm. In order to further reduce the reconstruction
rate of fake tracks, the radius of the cylinder of origin was reduced to
0.1~cm. The reconstruction was made faster by increasing the minimum $\pt$
cut to 0.175~GeV/$c$.

\begin{figure}
 \centering
  \includegraphics[width=0.49\linewidth]{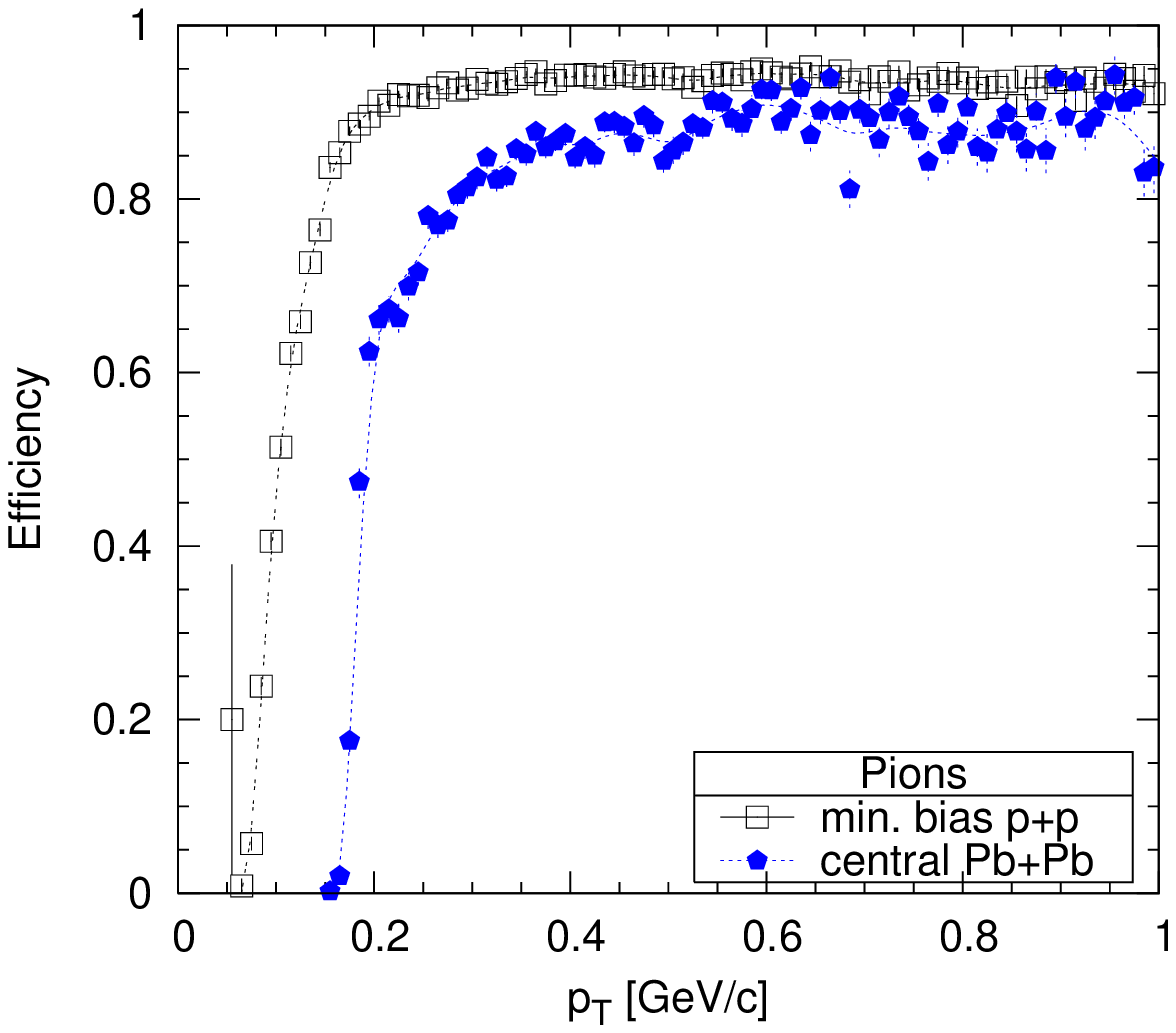}
  \includegraphics[width=0.49\linewidth]{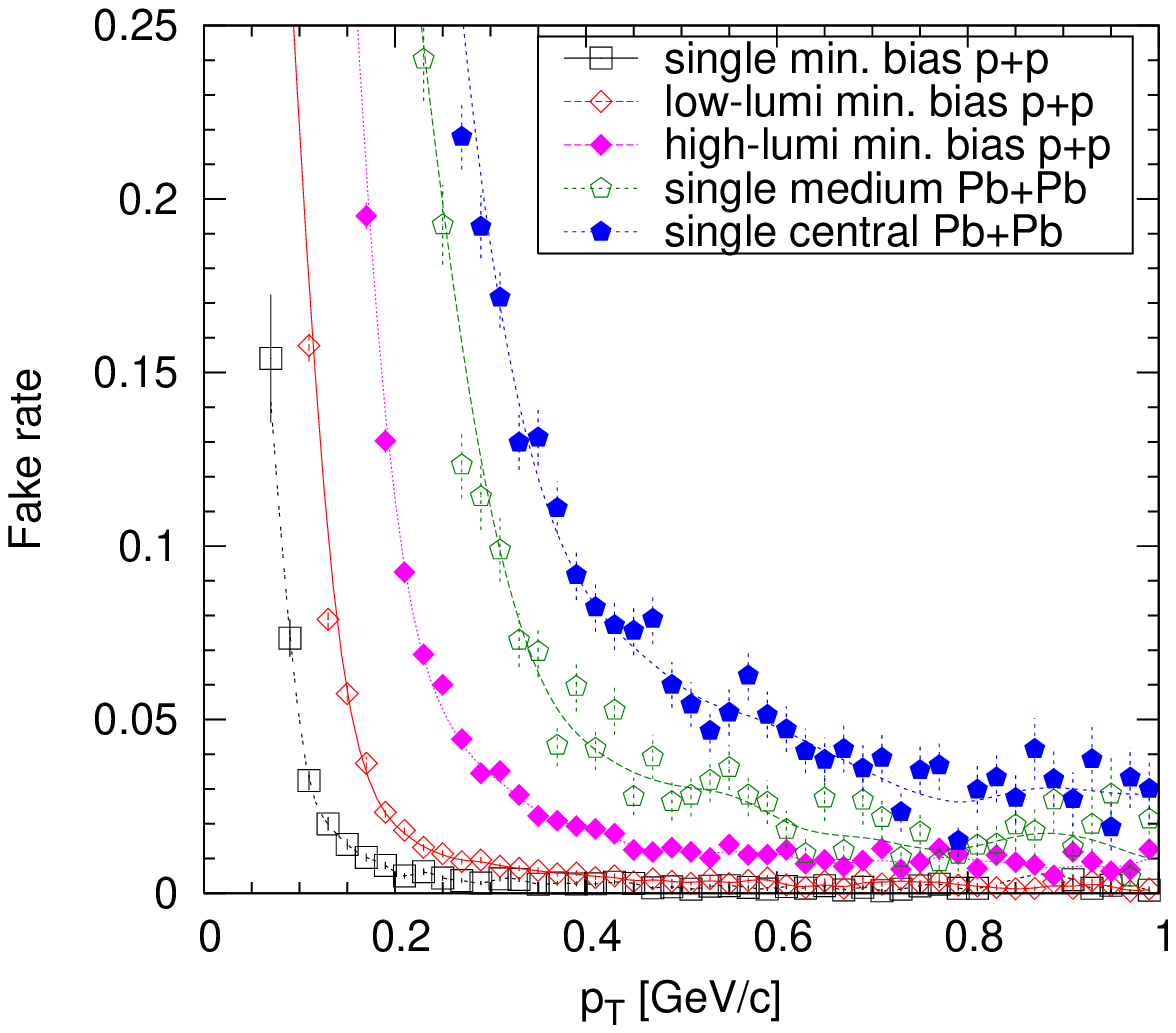}

 \caption{Left: Pion reconstruction efficiency as a function of $\pt$ for
tracks in the range $|\eta|<1$, for minimum bias pp events (squares) and for
central PbPb collisions (circles).  Right: Reconstruction rate of fake tracks
as a function of $\pt$, for tracks in the range $|\eta|<1$, for single, low
luminosity and high luminosity minimum bias pp events, and for central and
mid-central PbPb collisions.}

 \label{fig:comparison}
\end{figure}

The reconstruction efficiency is shown in Fig.~\ref{fig:comparison}-left, for
pions, as a function of $\pt$.  Above $\pt$ around 0.4~GeV/$c$, the pion
reconstruction efficiency in PbPb collisions is $\sim$~90\%, only 5\%
smaller than in pp collisions. Figure~\ref{fig:comparison}-right shows
that the reconstruction rate of fake tracks falls steeply with increasing
$\pt$.  It drops below 10\% for $\pt \sim 0.2$~GeV/$c$ in high-luminosity pp
collisions and for $\pt \sim 0.4$~GeV/$c$ in central PbPb collisions.

\section{Neutral hadron (V0) and (converted) photon identification}

It was shown in the previous section that silicon detectors can detect
charged particles with good position and momentum resolution. Some
weakly-decaying neutral particles (V0s) such as $\PKzS$, $\PgL$ and $\PagL$,
have a sizeable probability to decay far from the primary event vertex
($c\tau$ = 2.68 and 7.89 cm for $\PKzS$ and $\PgL$, respectively).  Likewise,
the silicon detectors can be used to reconstruct photons through their
conversion to $\mathrm{e^+e^-}$ pairs in the material of the beam-pipe,
silicon pixels and supports.

The analysis presented here only uses charged particles reconstructed from
pixel hit triplets, with a much wider cylinder of origin (3.0~cm).
Therefore, only neutral particles which decay up to the
first pixel barrel layer can be found. 
Considering their masses and $\pt$ distributions, 
about half of the produced 
$\PKzS$ and $\PgL$ particles satisfy this condition.


The search for V0 candidates reduces to the determination of the closest
point between two helices, as described in detail in
Ref.~\cite{Sikler:lowpt}.  The distribution of the distance between the decay
vertex and the beam-line ($r$) is shown in Fig.~\ref{fig:ph_k0s}-left. The
$r$ distributions for V0s show an exponential behaviour, steeper for $\PKzS$
than for $\PgL$, reflecting their different $c\tau$ values. The $r$
distribution for photons is completely different: the two peaks belong to the
inner and outer silicon wafers of the first pixel barrel layer.

\subsection{V0 results}

\begin{figure}
 \centering
  \includegraphics[width=0.49\linewidth]{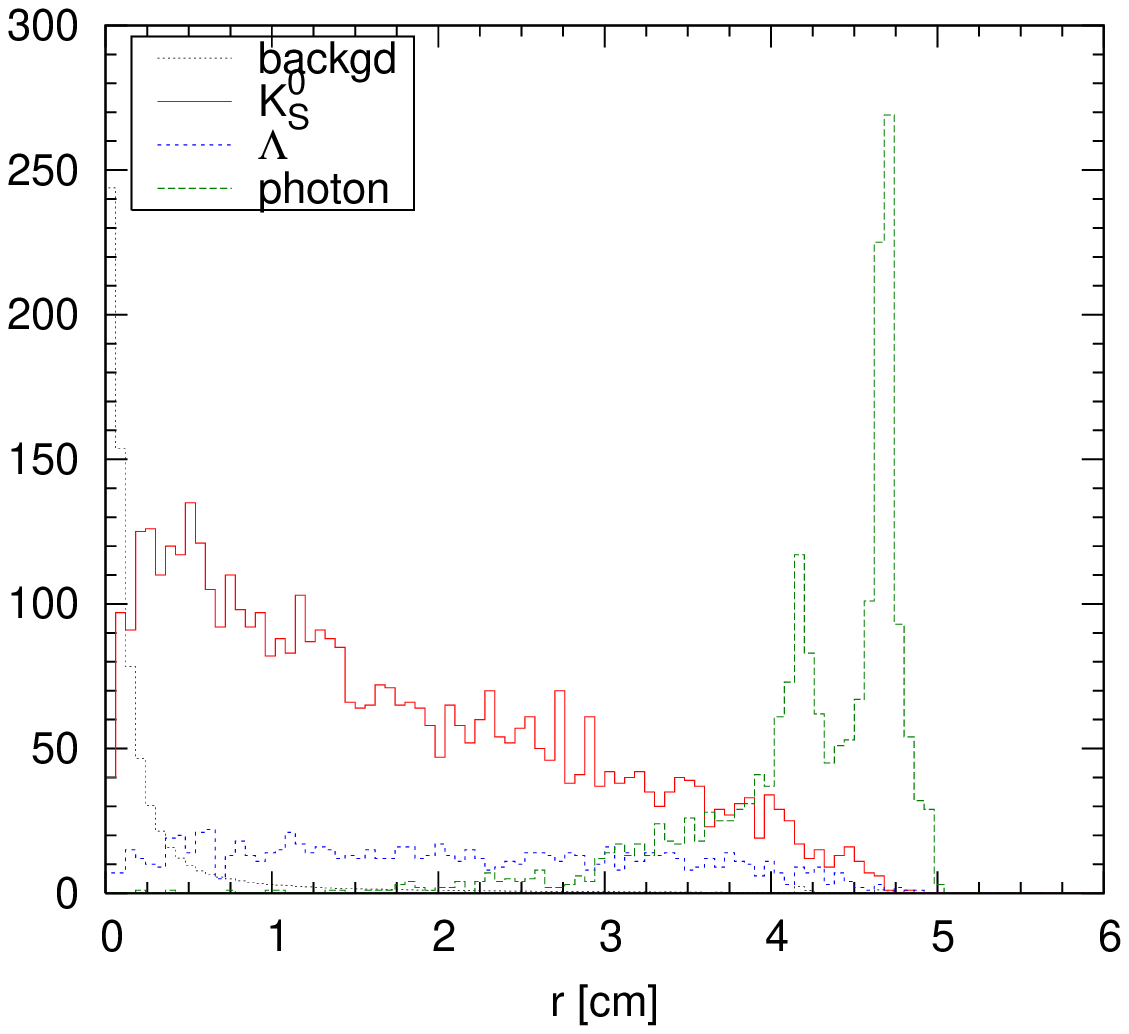}
  \includegraphics[width=0.49\linewidth]{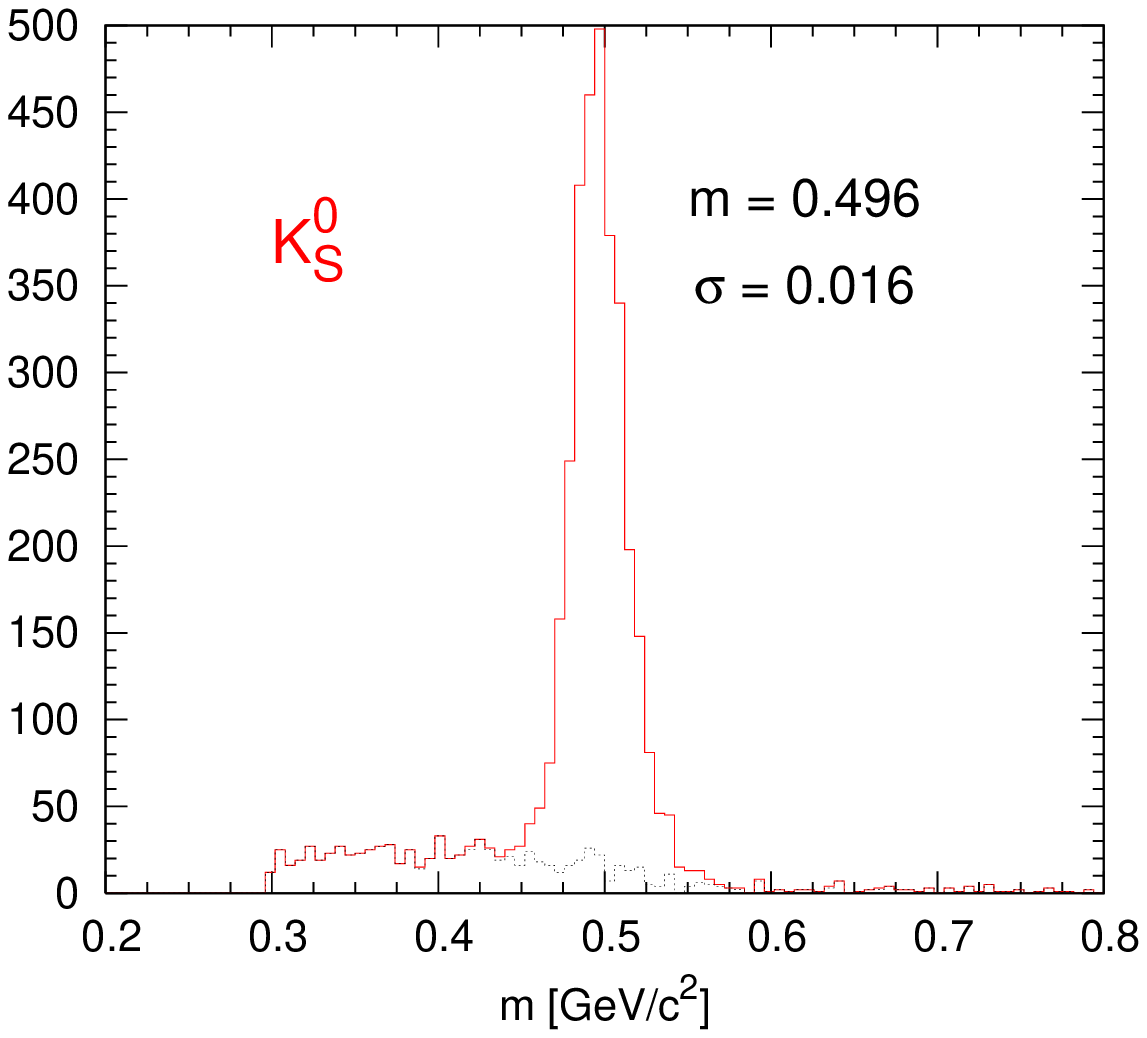}

 \caption{Left: Distribution of the distance ($r$)  between the ``decay
vertex'' and the beam-line for $\PKzS$, $\PgL$, converted photons and
background (fake particle pairs).  Right: Invariant mass distribution of
reconstructed $\PKzS\rightarrow \pi^+\pi^-$.  The mass distribution of the
background (fake pion pairs) is indicated by the dashed line. The
results of a Gaussian fit to the signal are given in units of GeV/$c^2$.}

 \label{fig:ph_k0s}
\end{figure}

These studies are based on 25~000 single minimum bias p+p events (Pythia
generator), reconstructed with the modified hit triplet finding.  The invariant
mass distribution of reconstructed $\PKzS\rightarrow \pi^+\pi^-$ decays is
shown in Fig.~\ref{fig:ph_k0s}-right.  The $\PKzS$ is reconstructed with a
resolution of 16~MeV/$c^2$, with an average mass of 0.496~GeV/$c^2$, in
agreement with the nominal mass value.  The $\PgL$ and $\PagL$ peaks (not
shown) are located at 1.114~GeV/$c^2$, with a resolution of 6~MeV/$c^2$.
Protons can be strongly enhanced by a cut on the truncated mean of their $\dedx$, removing
almost all the background.  In the case of single collisions or
low-luminosity pp running, the resonances can be exclusively identified.  For
high-luminosity pp running or PbPb collisions, the inclusive yield can still
be extracted, with a relatively small background.

\vspace{-0.09in}
\section{Conclusions}

With a modified hit triplet finding algorithm, the pixel detector can be
employed for the reconstruction of low $\pt$ charged hadrons in high
luminosity pp collisions, as well as in PbPb reactions. The acceptance of the
method extends down to 0.1, 0.2 and 0.3~GeV/$c$ in $\pt$ for pions, kaons and
protons, respectively. The fake track rate can be greatly reduced by using
the geometrical shape of the pixel clusters.  Weakly-decaying hadrons
($\PKzS$, $\PgL$ and $\PagL$) decaying before the first pixel layer can be
observed via their charged products.  Photons converting in the beam-pipe or
in the first pixel barrel layer are also detectable.

In summary, the CMS detector is able to provide good quality data on low
$\pt$ charged and neutral particle spectra and yields, thus contributing to
the soft hadronic physics program at the LHC.

\vspace{-0.09in}
\section*{Acknowledgements}

The author is thankful to David d'Enterria, Carlos Louren\c{c}o and other
members of the CMS Heavy Ion group for their valuable comments and
corrections to the text. This work was supported by the Hungarian Scientific
Research Fund (T 048898).

\end{document}